\documentclass[prl,twocolumn,amssymb,showpacs,floatfix,%
superscriptaddress]{revtex4}

\usepackage{graphicx}

\newcommand\apjl{Astrophys.~J. Lett.~}
\newcommand\apjs{Astrophys.~J. Suppl.~}
\newcommand\aap{Astron. Astrophys.~}
\newcommand\mnras{Mon. Not. R.~Astron. Soc.~}

\begin{document}
\title{Pulsar recoil by large-scale anisotropies in supernova explosions} 
\author{L. Scheck}
\affiliation{Max-Planck-Institut f\"ur Astrophysik, 
  Karl-Schwarzschild-Str. 1, D-85741 Garching, Germany}
\author{T. Plewa}
\affiliation{Center for Astrophysical Thermonuclear Flashes,
  University of Chicago, 5640 S.~Ellis Avenue, Chicago, IL 60637, USA}
\affiliation{Nicolaus Copernicus Astronomical Center, Bartycka 18,
  00716 Warsaw, Poland}
\author{H.-Th. Janka}
\affiliation{Max-Planck-Institut f\"ur Astrophysik, 
  Karl-Schwarzschild-Str. 1, D-85741 Garching, Germany}
\author{K. Kifonidis}
\affiliation{Max-Planck-Institut f\"ur Astrophysik,
  Karl-Schwarzschild-Str. 1, D-85741 Garching, Germany}
\author{E. M\"uller}
\affiliation{Max-Planck-Institut f\"ur Astrophysik,
  Karl-Schwarzschild-Str. 1, D-85741 Garching, Germany}

\date{\today}

\begin{abstract}
Assuming that the neutrino luminosity from the neutron 
star core is sufficiently high to drive supernova explosions
by the neutrino-heating mechanism, we show that low-mode
($l=1,2$) convection can develop from random seed 
perturbations behind the shock. A slow onset of
the explosion is crucial, requiring the core luminosity
to vary slowly with time, in contrast to the burst-like
exponential decay assumed in previous work. Gravitational and
hydrodynamic forces by the globally asymmetric supernova
ejecta were found to accelerate the remnant neutron star on
a timescale of more than a second to velocities above
500~km$\,$s$^{-1}$, in agreement with observed pulsar proper
motions.
\end{abstract}


\maketitle

Young pulsars are observed to have average space velocities 
of 200--500~km$\,$s$^{-1}$ with highest values above
1000~km$\,$s$^{-1}$ and still ambiguous hints for a double 
peak distribution~\cite{refskicks}. There is no
clear statistical correlation with the magnetic moment or 
rotation of the pulsar, although
in two cases (Vela and Crab) the direction of motion appears
to be aligned with the spin axis.

A connection of the pulsar motions with the supernova (SN) 
explosion is
suggested by neutron star (NS)$\,$--$\,$SN remnant associations
and by the properties of binary systems with one or both components 
being a NS. Natal kicks are required, e.g., by the spin-orbit
misalignment and high orbital eccentricities observed in some
binaries (for a review, see~\cite{lai01}).

Various mechanisms have been proposed to explain these kicks.
One possibility invokes asymmetric mass ejection during the 
SN~\cite{shk70}. This may be caused by large-scale
density inhomogeneities in the pre-collapse
core of the progenitor star~\cite{refsprog}
or convective instabilities in the neutrino-heated
layer behind the SN shock~\cite{refscon,jan94}. 
The kicks could also be a consequence of unequal momentum
fluxes in a jet and an anti-jet that might be linked to the start
of the SN explosion~\cite{SNjet} or to the NS formation~\cite{lai01}.

Alternatively or in
addition, anisotropic neutrino ($\nu$) emission from the nascent
(``proto-'') neutron star (PNS) might transfer the 
momentum~\cite{nukick}.
A ``neutrino rocket engine'' of the latter kind could
result from the magnetic field-strength dependence of 
$\nu$-matter interactions if extremely strong fields 
with hemispheric asymmetries build up in a PNS~\cite{Bkick}.
The $\nu$ transport also depends on the field
direction, e.g., through parity violating corrections of weak
interaction cross sections~\cite{Pkick} or due to resonant
flavor transitions~\cite{Osckick}.
In case of a significant dipole component such
effects can lead to kicks of a few 100~km$\,$s$^{-1}$ 
for magnetic fields in excess of 
$10^{15}\,$G~\cite{lai01}.


\begin{figure}[ht!]
  \begin{center} \leavevmode
   \includegraphics[width=0.85\columnwidth,clip=]{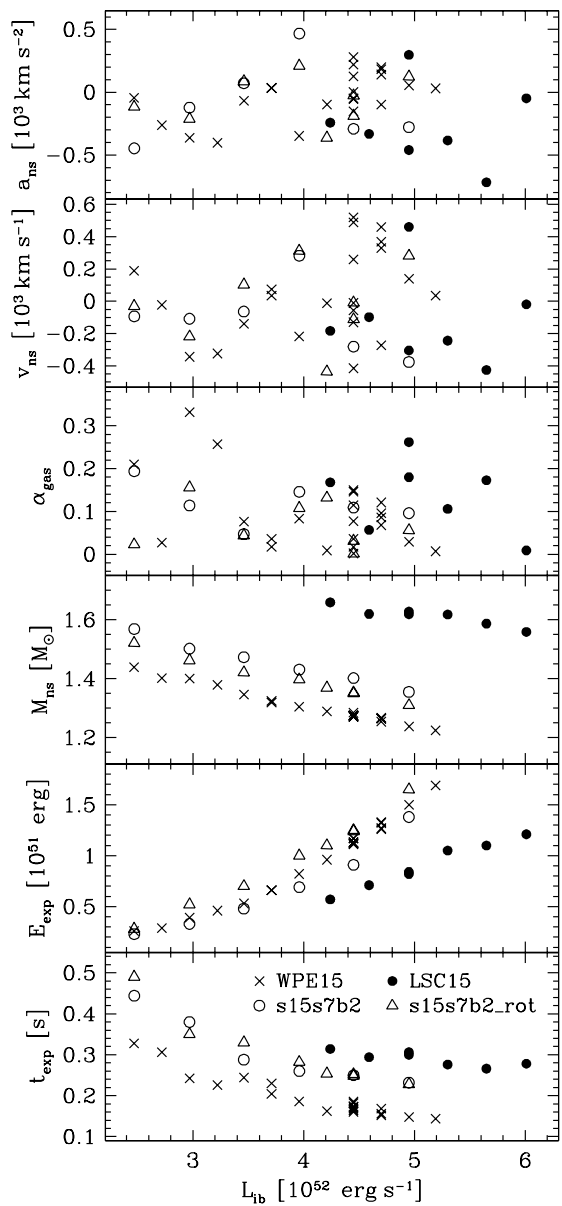}
   \vspace{0.25cm}
    \caption{Explosion timescale $t_{\mathrm{exp}}$ and,
             at one second after SN shock formation, explosion 
             energy $E_{\mathrm{exp}}$, NS baryonic mass 
             $M_{\mathrm{ns}}$, gas anisotropy parameter
             $\alpha_{\mathrm{gas}}$, NS recoil velocity
             $v_{\mathrm{ns}}$, and NS acceleration
             $a_{\mathrm{ns}}$ (from bottom to top) vs
             $\nu_e$ plus $\bar\nu_e$ luminosity,
             $L_{\mathrm{ib}}$, at the inner boundary.
             The symbols correspond to different models of 
             15$\,M_{\odot}$ stars, the triangles to a case 
             including rotation (see text for details).
     \label{fig:kick}}
   \end{center}
\end{figure}

\begin{figure*}[ht!]
   \begin{tabular*}{\textwidth}{l@{\extracolsep\fill}cr}
   \includegraphics[width=0.8\columnwidth,clip=]{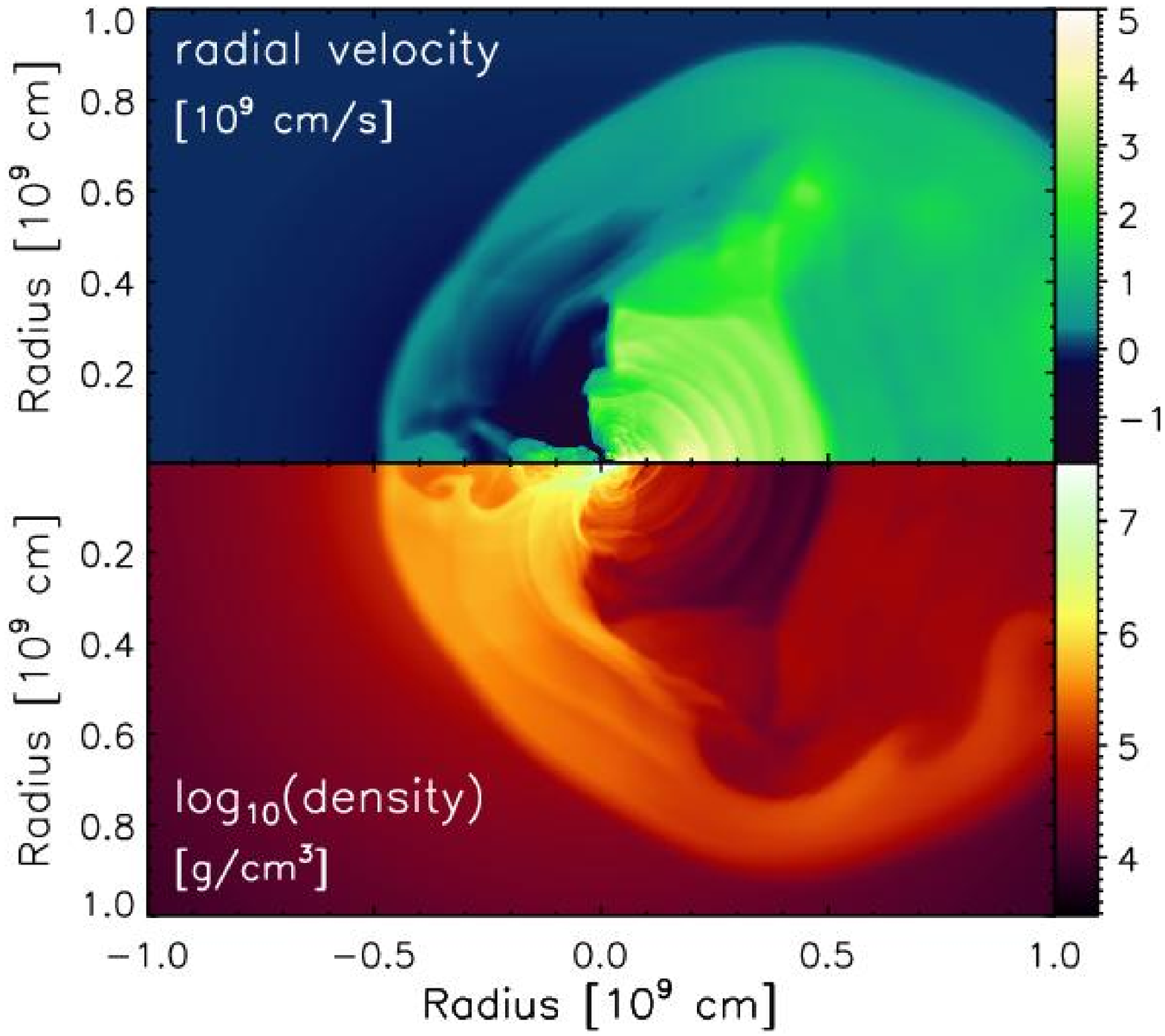} &
   \includegraphics[width=0.8\columnwidth,clip=]{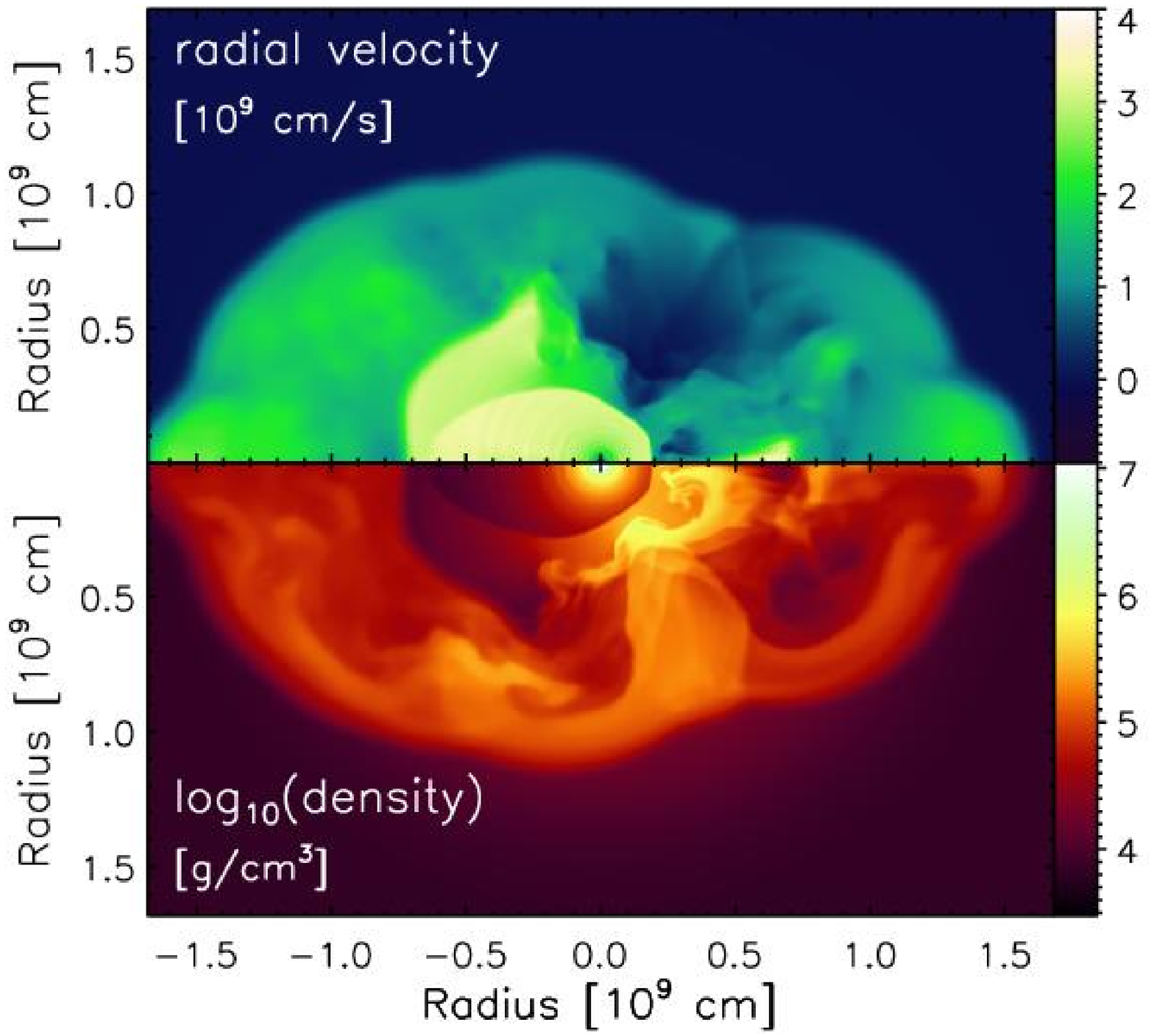} &
   \raisebox{6.5cm}{\parbox[t]{0.4\columnwidth}{ 
    \caption{Two models (based on progenitor
             WPE15) at 1$\,$s after SN shock formation.
             The PNS is at the coordinate center.
             The left plot (scale reduced to show more details)
             displays a case with $L_{\mathrm{ib}} = 
             2.97\times 10^{52}\,$erg$\,$s$^{-1}$,
             $E_{\mathrm{exp}} = 0.4\times 10^{51}\,$erg, and a
             recoil velocity of $v_{\mathrm{ns}} = -350$~km$\,$s$^{-1}$.
             The model on the right has
             $L_{\mathrm{ib}} = 4.45\times 10^{52}\,$erg$\,$s$^{-1}$,
             $E_{\mathrm{exp}} = 1.2\times 10^{51}\,$erg, and
             $v_{\mathrm{ns}} = +520$~km$\,$s$^{-1}$.
     \label{fig:models}}
   }} 
   \end{tabular*}
\end{figure*}

In this {\em Letter} we present new two-dimensional (2D)
calculations of hydrodynamic
instabilities during the onset of SN explosions which show that
global asymmetries and the PNS recoil can naturally
grow to a sufficient size without invoking artificial
initial conditions, extreme physical assumptions, or exotic 
$\nu$ physics. Our computations improve previous
ones~\cite{jan94} with respect to numerical resolution, a full
180$^{\mathrm{o}}$ lateral grid, and the treatment of $\nu$ transport,
extending them also in the computed evolution time and model set.

{\em Modeling concepts.}
We assume that the explosion is powered by $\nu$-energy deposition
between the PNS and the SN shock~\cite{bet85}. Although
the currently most elaborate numerical models are not able to confirm
the viability of this $\nu$-heating mechanism, it is still the 
best-studied and most promising way to explode massive 
stars~\cite{bur03}.

SN theory is currently hampered by
our incomplete knowledge of the nuclear physics and $\nu$
interactions in the dense matter 
inside the PNS. This implies uncertainties for self-consistent
models of the full problem, e.g.~with respect to the magnitude
of the $\nu$ fluxes emitted by the cooling PNS.
Therefore, we replace the shrinking,
high-density core of the PNS by a gravitating sphere whose
radius coincides with the contracting, impenetrable inner 
boundary of our computational grid.

At this boundary, number and energy fluxes of $\nu$ and
$\bar\nu$ of all three lepton flavors are
imposed with chosen initial values and time dependence.
In all simulations these boundary values
were taken to be {\em isotropic} 
and were kept constant for a chosen period of time
after shock formation (either 0.7$\,$s or 1$\,$s, 
within which 1/3 of the gravitational binding energy of the nascent 
NS was assumed to be radiated away in neutrinos).
The grid boundary is located somewhat below the
neutrinosphere of $\nu_e$. It has an optical depth which, 
for typical $\nu$ energies, rises from few initially
to several 100 at the end of the simulations. The use
of this inner boundary allows us to explore the response
of the collapsing SN core to different $\nu$ luminosities. Higher
values of the latter lead to larger energy deposition behind the 
SN shock and therefore to more powerful explosions. 
The dominant heating
reactions are $\nu_e$ and $\bar\nu_e$ absorption on free nucleons.

The ``lightbulb'' approximation employed previously~\cite{jan94}
ignored time retardation effects and did not take into account
radial variations of the fluxes due to $\nu$-matter interactions
in the cooling and heating layers between PNS and SN shock. 
To improve on that in the present work, we make use of the
zeroth moment of the Boltzmann transport equation in the form
$\ 
\partial_t L + \tilde c\, \partial_r L\,=\,
4\pi r^2 \tilde c\, Q_{\nu}^- - \kappa c L \,.\ 
$
Here $L = L(r,t)$ is the total $\nu$ number flux or $\nu$
luminosity, respectively, $Q_{\nu}^-$ the rate of $\nu$ loss 
by the stellar medium per unit volume,
$\kappa \equiv 4\pi r^2\tilde c\, Q_{\nu}^+/(Lc)$ 
the corresponding absorptivity, $c$ the speed of light, and
$\tilde c$ the ``effective speed'' of $\nu$ propagation, which is
governed by diffusion at high densities and reaches $c$ at 
large radii. The integration for $L$ in radius $r$ and time $t$
(lateral flux components are ignored) can be done analytically
when $Q_{\nu}^-$, $\kappa$ and $\tilde c$ are
assumed to be constant within the cells of the 
numerical grid (consistent with this, the above equation was
derived by employing $\partial_t \tilde c = 0$). Instead of
determining $\tilde c$ from the solution of the Boltzmann equation,
we use an analytic representation in terms of the optical depth
that was obtained from fitting results of detailed $\nu$
transport in the outer layers of the SN core.
Neutrino-matter interactions via charged-current processes 
with nucleons, thermal pair creation, and
scattering off nuclei, $n$, $p$, $e^-$, and $e^+$ are evaluated by
adopting Fermi-Dirac $\nu$ spectra with a
temperature determined by the ratio between $\nu$ energy and
number density. The chemical potentials of the spectra are taken
to be equal to the equilibrium values at high optical
depths and approach values near zero outside of the 
neutrinospheres.

This approximate $\nu$ transport retains the
hyperbolic character of the transport problem, conserves lepton 
number and energy globally, and reproduces basic properties
of accurate Boltzmann transport calculations despite of radical
simplifications. It is coupled to our 2D hydrodynamics code
by operator-splitting with a predictor-corrector step.

Our hydrodynamics code and equation of state
are described in Ref.~\cite{kif03}. We typically use
400 geometrically spaced radial zones and one degree lateral
resolution. The $\nu$ transport is solved in each angular bin
separately. The 2D simulations were started some milliseconds 
after SN shock formation from detailed core-collapse models
with fluid velocities randomly perturbed with an amplitude
of typically 0.1\%.

{\em Results.}
Figure~\ref{fig:kick} shows four sequences of runs that
followed the post-bounce evolution of different 
15$\,M_{\odot}$ progenitors for one second. 
The crosses mark results based on the use of 
Model~WPE15~\cite{woo88}, the solid dots of
Model~LSC15~\cite{lim00},
the open circles of Model~s15s7b2~\cite{woo95},
and the triangles of a model with a structure like
the latter but including rotation~\cite{bur03}.
Starting with a rotation period of 12$\,$s in the
iron core~\cite{woo02}, 
the post-collapse core spins differentially within 10--20~ms
and speeds up as it contracts. 
The rotation period in the $\nu$-heating layer varies
between some $10\,$ms and several 100$\,$ms~\cite{bur03}. 

Model~LSC15, in particular,
differs from the others by having significantly
higher densities at the edge of the iron core and in 
the silicon shell (at a time when the cores have evolved 
to the same central density). This delays the start time,
$t_{\mathrm{exp}}$, of the explosion, reduces the explosion
energy, $E_{\mathrm{exp}}$, and leads to a larger NS baryonic mass,
$M_{\mathrm{ns}}$, for a given value of the boundary
luminosity, $L_{\mathrm{ib}}$, of $\nu_e$ plus $\bar\nu_e$
(Fig.~\ref{fig:kick}). $E_{\mathrm{exp}}$ is defined
as the total energy (internal plus kinetic plus gravitational)
of the SN ejecta, integrated over
all matter where the sum of the corresponding specific energies
is positive, $t_{\mathrm{exp}}$ is the post-bounce time when
$E_{\mathrm{exp}}$ reaches $10^{49}\,$erg, and 
$M_{\mathrm{ns}}$ is the gas mass with densities
above $10^{11}\,$g$\,$cm$^{-3}$ plus the central
point mass at one second.

\begin{figure}[ht!]
  \begin{center} \leavevmode
   \includegraphics[width=0.80\columnwidth,clip=]{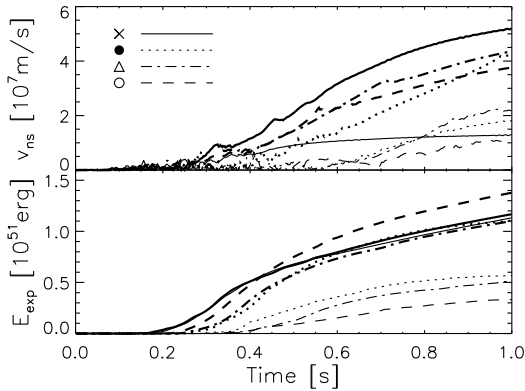}
   \vspace{0.25cm}
    \caption{Explosion energies (bottom) and NS velocities
             vs time for some simulations, showing large
             acceleration for cases with high $v_{\mathrm{ns}}$ 
             (thick lines) even at 1$\,$s after bounce. The symbols 
             and line styles refer to the different model
             sequences of Fig.~\ref{fig:kick}.
     \label{fig:tevol}}
   \end{center}
\end{figure}

Figure~\ref{fig:kick} reveals the generic trend that a higher
luminosity $L_{\mathrm{ib}}$ from the NS core causes the 
explosion to develop faster and to become more energetic. 
Because the period of mass accretion by the PNS is reduced,
this implies a smaller NS mass. 
The time until the revived bounce-shock reaches
a certain radius (correlated with
$t_{\mathrm{exp}}$) depends
sensitively on the progenitor structure and the
core $\nu$ luminosity. Rotation systematically
increases the explosion energy by 20--50\%, 
because centrifugal forces delay
matter from being accreted onto the PNS, and thus keep it in the 
$\nu$-heating region to accumulate energy by $\nu_e$ and
$\bar\nu_e$ absorption. This is basically in agreement with
Ref.~\cite{bur03}, where rotation was found to stabilize 
the standing accretion shock at a larger radius.

The layer between PNS and SN shock is convectively unstable
according to the Ledoux-criterion because of a negative
entropy gradient established by $\nu$-energy deposition.
Within $\sim$$50\,$ms after shock formation, convection
sets in, supporting the start of the 
explosion~\cite{refscon,jan94}. Initially the convective
cells are small, but they begin to merge to larger entities.
Three-dimensional (3D) simulations agree with this
2D result~\cite{fry02}. In case of rapid shock acceleration
convection freezes, and small structures
characterize the flow pattern until late times. However, when 
the stagnant 
shock expands slowly, small cells have time to merge to very
large buoyant bubbles, separated by
only a few narrow, supersonic downflows which carry low-entropy
matter from the shock to the PNS surface. Moreover,
global pulsations can develop with a 
dominance of low-order ($l = 1,2$) modes. 
Consequently, the density distribution becomes highly 
aspherical and the explosion breaks out with a very large
hemispheric or polar-to-equatorial asymmetry and corresponding 
shock deformation. In some cases a single long-lasting
accretion funnel was found to persist for a second or even
longer (Fig.~\ref{fig:models}). We emphasize that such
structures did not preferentially occur along the polar 
\hbox{($z$-)} direction
of our spherical grid (where a coordinate singularity exists),
but developed in arbitrary orientations.

The development of a stable, volume-filling $l=1$ mode 
was proposed before~\cite{lmode95}. It is supported by
analytic arguments for thermal instabilities in fluid 
spheres~\cite{cha81} and is also observed in 3D
hydrodynamic simulations of pulsating, convective red giant
stars~\cite{wood03}. Determining the duration of such a
phenomenon in the time-dependent environment of an exploding SN
requires numerical modeling. The coherent, low-order
oscillations of the fluid beneath the shock in our simulations
look similar to the recently discovered 
non-radial instabilities in adiabatic flow behind standing
accretion shocks~\cite{blo03}, which can be understood in
terms of a ``vortical-acoustic cycle''~\cite{fog02}. Doing
numerical experiments we found that this phenomenon might indeed
play a role, although Ledoux-instability due to $\nu$-heating
clearly starts the convective activity, and $\nu$-cooling around 
the neutrinosphere seems to damp the energetic amplification of 
the feedback cycle between turbulence and pressure waves.

Anisotropic mass ejection can be associated with a 
high linear 
momentum taken up by the compact remnant. The corresponding 
asymmetry of the explosion is expressed by the parameter
$\alpha_{\mathrm{gas}} \equiv |P_{z,\mathrm{gas}}|/P_{\mathrm{gas}}
\equiv |\int {\mathrm{d}}m\,v_z|/\int {\mathrm{d}}m\,|\vec v|$
where the integrals are performed over the ejecta mass and 
$P_{z,\mathrm{gas}}$ is the gas momentum along the $z$-direction of
the 2D grid. We find values for
$\alpha_{\mathrm{gas}}$ up to 0.33 (Fig.~\ref{fig:kick}).
The NS recoil velocities,
$v_{\mathrm{ns}} = \alpha_{\mathrm{gas}}P_{\mathrm{gas}}/
M_{\mathrm{ns}}$, can be close to zero but also more than 
500~km$\,$s$^{-1}$. In some cases the
acceleration, $a_{\mathrm{ns}}$, continues on a
high level beyond the 1$\,$s of computed evolution
(Figs.~\ref{fig:kick}, \ref{fig:tevol}) and significantly larger
terminal velocities can be expected. It is mediated by 
the long-range gravitational
force between the asymmetrically distributed ejecta and the
PNS. Hydrodynamic forces play a role only as long as downflows
reach the PNS, and anisotropic $\nu$ emission
contributes insignificantly. There is no correlation
of $v_{\mathrm{ns}}$ with the progenitor model. Rotation does not
inhibit large kicks. The final recoil velocity depends
stochastically on the initial seed perturbation 
and the highly nonlinear growth of the convective structures.
There is also no obvious correlation with the explosion energy.
Fig.~\ref{fig:tevol} (by the thin and thick solid lines) and 
Fig.~\ref{fig:kick} (coinciding points) demonstrate that nearly
the same $E_{\mathrm{exp}}$ can be associated with large or 
small $v_{\mathrm{ns}}$.

{\em Conclusions.}
We have shown that globally anisotropic mass
ejection and NS acceleration can result from convective 
overturn and low-order oscillations of the $\nu$-heated layer
in a SN core. Low-mode convection turned out to develop
from random seed perturbations in case of a slow
onset of the explosion which gives the convective structures
time to merge. This was disfavored by our previous choice of
a strongly time-dependent, exponentially decaying core $\nu$
luminosity in Refs.~\cite{jan94,kif03} but is possible with the
less burst-like (because constant) $L_{\mathrm{ib}}$ assumed in 
this work.
Our models suggest a consistent
picture in which $\nu$-energy deposition can be responsible
for the SN explosion, for pulsar kicks, and for
global asymmetries observed in many supernovae.
The simulations need to be continued to later times to allow for
quantitative conclusions on the morphological properties of the
ejecta as, e.g., inferred from polarization measurements.
Statistical information about the distribution
of intrinsic pulsar velocities requires 3D simulations,
a better fundamental understanding of the explosion mechanism,
and a large sample of simulations for progenitor stars with 
different masses and rotation rates. The discussed kick mechanism
does not enforce a strict alignment of pulsar spin and space
velocity, although the rotation axis of a star defines a
naturally preferred direction, which might disfavor 
large misalignments. 

We thank S.W.~Bruenn and M.~Rampp for 
post-bounce core-collapse models and M.~Limongi and
S.~Woosley for their progenitor models.
Support by the Sonderforschungsbereich
375 on ``Astroparticle Physics'' of the Deutsche
Forschungsgemeinschaft is acknowledged.
The simulations were done at the Rechenzentrum Garching and
the Interdisciplinary Centre for Computational Modelling in
Warsaw.

\end{document}